\documentclass[conference]{IEEEtran}
\usepackage{amsfonts}
\usepackage{amsmath,amscd}
\usepackage{amssymb}
\usepackage{graphicx}%
\usepackage{color}
\usepackage{tikz}	
\usetikzlibrary{backgrounds,fit,decorations.pathreplacing}  % TikZ libraries

\newtheorem{theorem}{Theorem}

 \DeclareMathOperator{\tr}{Tr}

\newcommand{\ket}[1]{| #1 \rangle}
\newcommand{\proj}[1]{| #1 \rangle\!\langle #1 |}

\newcommand{\be}{{\mathbf e}}

\def\cD{{\cal D}}        \def\cE{{\cal E}}

        \def\cN{{\cal N}}

\def\0{{\mathbf{0}}}
\def\1{{\mathbf{1}}}
\def\2{{\mathbf{2}}}
\def\3{{\mathbf{3}}}
\def\4{{\mathbf{4}}}
\def\5{{\mathbf{5}}}
\def\6{{\mathbf{6}}}

\def\7{{\mathbf{7}}}
\def\8{{\mathbf{8}}}
\def\9{{\mathbf{9}}}

\def\be{\begin{equation}}
\def\ee{\end{equation}}
\def\bea{\begin{eqnarray}}
\def\eea{\end{eqnarray}}

\begin{document}

\title{Fully Quantum Source Compression with a Quantum Helper}
\author{
\IEEEauthorblockN{Min-Hsiu Hsieh}
\IEEEauthorblockA{
University of Technology, Sydney\\
Email: Min-Hsiu.Hsieh@uts.edu.au}
\and
\IEEEauthorblockN{Shun Watanabe}
\IEEEauthorblockA{
Tokyo University of Agriculture and Technology \\
Email: shunwata@cc.tuat.ac.jp }}

\maketitle

\begin{abstract}
We study source compression with a helper in the fully quantum regime, extending our earlier result on classical source compression with a quantum helper [\textit{arXiv:1501.04366, 2015}].  We characterise the quantum resources involved in this problem, and derive a single-letter expression of the achievable rate region when entanglement assistance is available. The direct coding proof is based on a combination of two fundamental protocols, namely the quantum state merging protocol and the quantum reverse Shannon theorem (QRST). This result demonstrates an unexpected connection between distributed source compression with the QRST protocol, a quantum  protocol that consumes noiseless resources to simulate a noisy quantum channel.
\end{abstract}

%%%%%%%%%%%%%%%%%%%%%%%%%%%%%%%%%%%%
%  INTRODUCTION
%%%%%%%%%%%%%%%%%%%%%%%%%%%%%%%%%%%%

\section{Introduction}

Quantum Shannon theory has been extensively studied in the past decade \cite{Wilde13}. It not only generalises Shannon's source and channel coding theorems \cite{Shannon:1948wk}, but also includes scenarios that do not have classical correspondences due to the existence of entanglement. Unlike classical correlations, i.e.~the common randomness, quantum entanglement fundamentally changes an information-processing task whenever it is involved in the protocol. The study of bipartite information-processing tasks reached its peak in Ref.~\cite{Hsieh:2010gd}, where tradeoffs between classical communication, quantum communication, and entanglement for major protocols that either involve a quantum channel or a quantum source in the Shannon-theoretic setting are given. In the quantum multipartite information-processing tasks, the capacity regions become much more complicated and often require regularisation even with entanglement assistance \cite{HDW08,DHL10}.  Moreover, a quantum network information theory is still in its infancy \cite{Savov12}. 

Quantum information theorists have developed a resource theory that can efficiently describe an information-processing task and largely simply the development of coding theorems for complicated protocols that involve multiple valuable resources \cite{Devetak:2004go, Devetak:2008eb, Datta:2011vc}.  The extension of classical Slepian-Wolf coding theorem to the quantum regime has obtained great success in quantum Shannon theory since the fully quantum Slepian-Wolf (FQSW) protocol turns out to sit on the top of the hierarchy chains of family protocols \cite{Devetak:2004go, Devetak:2008eb, Datta:2011vc, Abeyesinghe:2009ej}. From the FQSW protocol, coding theorems for many other fundamental protocols can be trivially obtained by combining it with teleportation, superdense coding, or entanglement distribution \cite{Hsieh:2010gd, Devetak:2004go, Devetak:2008eb}.  It also includes the quantum source coding pioneered by Schumacher \cite{Schumacher:1995dg}, followed by many others \cite{Jozsa:1994ea, HM02}. 

In the task of classical source coding with side information, there is no advantage if Alice has additional side information regarding the compressed message $X$. Moreover, the shared common randomness between Alice and Bob is also of no help. Therefore, the classical Slepian-Wolf theorem is the most general theorem for source coding with side information. In sharp contrast, additional quantum side information at the encoder and/or the decoder changes the problem \cite{Yard:2009ib}. Besides, pre-shared entanglement between Alice and Bob, the quantum analog of shared common randomness, also proves to be useful \cite{Yard:2009ib}. The corresponding protocol, the \textit{state redistribution}, characterises the cost for Alice who owns systems $A$ and $C$ to redistribute part of her system $C$ to Bob when originally Bob has system $B$ and the inaccessible referee has system $R$ of a pure state $\ket{\psi_{ABCR}}$. The communication cost is determined by the conditional mutual information $\frac{1}{2}I(C;R|B)_\psi$, giving the first operational meaning for the quantum conditional mutual information. 

What if the quantum side information is only observed by a distant helper in the problem of source coding? Will an answer to this problem further deepen our understanding of the power of quantum resources? In our earlier study \cite{HW15}, we provided a partial solution by extending the classical source coding problem with a classical helper \cite{Wyner:1975iv, Ahlswede:1975ea}
to the classical source coding problem with a quantum helper.
We consider a classical-quantum scenario, where the sender Alice has a classical source while the helper observes a correlated quantum system with the source and can only communicate with the decoder through a classical channel. We derived a single-letter characterization of the achievable rate region, where the direct part of our result is proved via the measurement compression theory by Winter \cite{Winter:2004uk, Wilde:2012iq}; such an approach is a reminiscence of the approach 
taken in \cite{Watanabe:2013ea, Watanabe:2015ea} to derive a non-asymptotic bound on
the classical distributed source coding problem with a classical  helper.
Our result reveals that a helper's strategy that separately conducts a measurement and a compression is sub-optimal, and the measurement compression is fundamentally needed to achieve the optimal rate region.

In this paper, we extend the classical distributed source coding problem \cite{Wyner:1975iv, Ahlswede:1975ea} and its classical-quantum generalisation \cite{HW15} to the fully quantum version; namely compression of a quantum source with the help of a quantum server. Moreover, we consider a general setting where entanglement assistance between sender-decoder and helper-decoder is available. This answers the open question raised in \cite{HW15}. Our direct coding proof combines two fundamental quantum protocols; the state merging protocol \cite{Horodecki:2005fv, Horodecki:2006hl} and the quantum reverse Shannon theorem  \cite{Bennett:2014il}.

\textit{Notations.} Various entropic quantities will be used in the paper. The von Neumann entropy of a quantum state $\rho_A$, where the subscript $A$ represents the quantum state is held by A(lice), is $H(A)_\rho=-\tr(\rho_A\log\rho_A) $. The conditional von Neumann entropy of system $A$ conditioned on $B$ of a bipartite state $\rho_{AB}$ is $H(A|B)_\rho = H(AB)_\rho - H(B)_\rho$. The quantum mutual information between two systems $A$ and $B$ of $\rho_{AB}$ is $I(A;B)_\rho= H(A)_\rho+H(B)_\rho-H(AB)_\rho$. The conditional quantum mutual information $I(A;B|C)_\rho = I(A;BC)_\rho-I(A;C)_\rho$.

Before introducing quantum protocols that will be needed in our main result, we will first review the language of \textit{Resource Inequality (RI)} \cite{Devetak:2008eb, Datta:2011vc}. The RIs are a concise way of describing interconversion of  resources in an information-processing task. Denote by $[qq]$ and $[q\to q]$ an ebit (maximally entangled pairs of qubits) and a noiseless qubit channel, respectively. Then a quantum channel $\cN$ that can faithfully transmit $Q(\cN)$ qubits per channel use with an unlimited amount of entanglement assistance can be  symbolically represented as
\[
\langle \cN \rangle +\infty [qq] \geq Q(\cN) [q\to q],
\]
where $\langle \cN \rangle$ is an asymptotic noisy resource that corresponds to many independent uses, i.e. $\cN^{\otimes n}$. Schumacher's noiseless source compression \cite{Schumacher:1995dg} can be similarly expressed
\[
H(B)_\rho [q\to q] \geq  \langle \rho_B\rangle,
\] 
which means that a rate of $H(B)_\rho$ noiseless qubits \emph{asymptotically} is sufficient to represent the noisy quantum source $\rho_B$. 

Sometimes, the RI only applies to the relative resource, $\langle \cN:\rho \rangle$, which means that the asymptotic accuracy is achieved only when $n$ uses of $\cN$ are fed an input of the form $\rho^{\otimes n}$. For detailed treatment of combining two RIs and rules of cancellation of quantum resources, see Ref.~\cite{Devetak:2008eb}.

\section{Relevant quantum protocols}

Given a bipartite state $\rho_{AB}$ whose purification is $\ket{\psi_{ABR}}$, the state merging protocol \cite{Horodecki:2005fv, Horodecki:2006hl, Dupuis:2014jz} is the information-processing task of distributing $A$-part of the system that originally belongs to Alice to the distant Bob without altering the joint state. Moreover, Alice and Bob have access to pre-shared entanglement and their goal is to minimise the number of EPR pairs consumed during the protocol. The state merging can be efficiently expressed as the following RI:
\begin{equation}
\langle \psi_{A|B|R}\rangle +I(A;R)_\psi[c\to c] +H(A|B)_\psi[qq] \geq \langle \psi_{|AB|R}\rangle
\end{equation}
where the notation $\psi_{A|B|R}$ denotes the state is originally shared between three distant parties Alice, Bob, and Eve, while $\psi_{|AB|R}$ means that the system $A$ is now together with system $B$. This protocol involves classical communication; however, for the purpose of this paper, quantum resources are much more valuable and classical communication is considered to be free. As a result, the state merging protocol either consumes EPR pairs with rate $H(A|B)_\psi$ when this quantity is positive, or generates $H(A|B)_\psi$ rate of EPR pairs for later uses, if $H(A|B)_\psi$ is negative, after the transmission of the system $A$ to $B$. 

The state merging protocol gives the first operational interpretation to the conditional von Neumann entropy. More importantly, it provides an answer to the long-standing puzzle---the conditional von Neumann entropy could be negative, a situation that has no classical correspondence.

The fully quantum Slepian-Wolf (FQSW) protocol \cite{Abeyesinghe:2009ej, Datta:2011vc} can be considered as the coherent version of the state merging protocol. It can be described as
\begin{equation}
\langle \psi_{A|B|R}\rangle + \frac{1}{2} I(A;R)_\psi [q\to q] \geq \frac{1}{2} I(A;B)_\psi [qq] +\langle \psi_{|AB|R}\rangle.
\end{equation} 
It is a simple exercise to show, via the resource inequalities, that the state merging protocol can be obtained by combining teleportation with the FQSW protocol \cite{Abeyesinghe:2009ej, Hsieh:2010gd}. Moreover, the FQSW protocol can be transformed into the a version of the quantum reverse Shannon theorems (QRST) that involves entanglement assistance \cite{Abeyesinghe:2009ej}. 

The quantum reverse Shannon theorem (QRST) addresses a fundamental task that asks, given a quantum channel $\cN$, how much quantum communication is required from Alice to Bob so that the channel $\cN$ can be simulated. There are variants of the QRSTs depending on whether entanglement or feedback is allowed in the simulation (see \cite[Theorem 3]{Bennett:2014il}).  The QRST protocol has become a powerful tool in quantum information theory. It can be used to establish a strong converse to the entanglement-assisted capacity theorem. Moreover, it can also be used to establish quantum rate distortion theorems  \cite{Datta:2013ur, Wilde:2013hp, Datta:2013jk}.

In this paper, we will use the QRST with entanglement assistance. 
\begin{theorem}[Quantum Reverse Shannon Theorem]\label{thm_QRST}
 Let $\cN$ be a quantum channel from $A$ to $B$ so that its isometry $U^\cN_{A\to BE}$ results in the following tripartite state when inputting $\rho_A$:
\[
\ket{\psi_{RBE}}=U^\cN_{A\to BE} \ket{\psi^\rho_{RA}},\]
where $\tr_R\proj{\psi_{RA}^\rho}=\rho_A$. Then with sufficient amount of pre-shared entanglement, the channel $\cN$ with input $\rho_A$ can be simulated with quantum communication rate $\frac{1}{2}I(R;B)_\psi$:
\begin{equation}\label{eq:QRST}
\frac{1}{2} I(R;B)_\psi [q\to q]+ \frac{1}{2} I(E;B)_\psi[qq]\geq \langle \cN:\rho_A\rangle. 
\end{equation} 

\end{theorem}

\section{Main Result}

\begin{figure}
\centerline{
    \begin{tikzpicture}[scale=1][very thick]
    \fontsize{10pt}{1} %%{font size}{line space}
    \tikzstyle{halfnode} = [draw,fill=white,shape= underline,minimum size=1.0em]
    \tikzstyle{checknode} = [draw,fill=blue!10,shape= rectangle,minimum height=3.5em, minimum width=2em]
    \tikzstyle{checknode2} = [draw,fill=blue!10,shape= rectangle,minimum height=8em, minimum width=2em]
    \tikzstyle{variablenode} = [draw,fill=white, shape=circle,minimum size=0.8em]
%    \node (p1) at (4.5,3) {$R^n$} ;
    \node (e1) at (-2,1) {$\rho_{AB}^{\otimes n}$} ;
    \node (EPR1) at (-2,2.5) {$\Phi_{T_AT_A'}$} ;
    \node (EPR2) at (-2,-0.5) {$\Phi_{T_BT_B'}$} ;
    \node (p3) at (4.5,1) {$\widehat{A}^n$} ;
    \node (p4) at (4.5,2.0) {$C_1$} ;
    \node (p5) at (4.5,0.4) {$\widehat{L}$} ;
    \node (p6) at (4.5,-0.1) {$\widehat{T}_{B}'$} ;
    \node (w2) at (1.5,1.25) {$M$} ;
    \node (w3) at (0.8,2.2) {$A_1$} ;
    \node (l2) at (1.5,0.5) {$L$} ;
    \node (s2) at (-0.7,1.3) {$A^n$} ;
    \node (e2) at (-0.7,0.7) {$B^n$} ;
    \node (s3) at (-0.7,2.2) {$T_A$} ;
    \node (e3) at (-0.7,-0.2) {$T_B$} ;
    \node (s4) at (2.3,2.25) {$T_A'$} ;
    \node (e4) at (2.3,-0.35) {$T_B'$} ;
 % draw lines between nodes
% \draw[->] (cn1) -- (cn2 |- xline);
%     \draw  (cn1) --  (cn2;%(cn1) -- (label6);
%     \draw  (-1,3)  -- (p1) (-1.5, 1.0) -- (-1,3); %R^n
     \draw  (-1.5,1)-- (-1,1.5) (-1.5,1)-- (-1,0.5) (-1,0.5) -- (0,0.5) (-1,1.5) -- (0,1.5); % \rho_AB
     \draw  (-1.5,-0.5)-- (-1,0) (-1.5,-0.5)-- (-1,-1.0) (-1,0) -- (0,0) (-1,-1) -- (1.5,-1) (1.5,-1)--(2,-0.1); % EPR below
     \draw  (-1.5,2.5)-- (-1,2) (-1.5,2.5)-- (-1,3.0) (-1,2.0) -- (0,2.0) (-1,3.0) -- (1.5,3.0) (1.5,3.0)-- (2, 2.0); % EPR above
     \draw (3.3,1) --++ (p3) (3.3,2.0) --++ (p4) (3.3,0.4) --++ (p5) (3.3,-0.1) --++ (p6); % D output
     \draw (0.3,0.25) --(3,0.25) (2,-0.1) --(3,-0.1) ; % E_B output
     \draw  (0.3,1.5)--(3,1.5) (0.3,1.45)--(3,1.45) (0.3, 2.0) -- (1,2.0) (2, 2.0) -- (3,2.0); % E_A output
    \node[checknode] (cn1) at (0,1.75) {${\cal E}_{A}$};
    \node[checknode2] (cn2) at (3,1) {${\cal D}$};
    \node[checknode] (cn3) at (0,0.25) {${\cal E}_{B}$};  
   \end{tikzpicture}
 }

  \caption{
Fully Quantum Source Compression with a Quantum Helper.
  }\label{fig:QSCQH}
\end{figure}
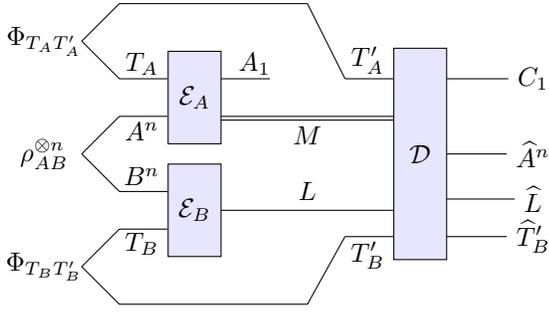

As shown in Figure~\ref{fig:QSCQH}, the protocol for fully quantum source coding with a quantum helper involves two senders, Alice and Bob, and one receiver, Charlie. Initially Alice and Bob hold $n$ copies of a  bipartite quantum state $\rho_{AB}$, where Alice holds quantum systems $A^n:=A_1\cdots A_n$ while Bob (being a quantum helper) holds quantum systems ${B^n}=B_1\cdots B_n$. Moreover, there are pre-shared entangled states $\ket{\Phi_{T_AT_A'}} $ between Alice and Charlie, and pre-shared entangled states $\ket{\Phi_{T_BT_B'}} $ between Bob and Charlie. The goal is for the decoder Charlie to faithfully recover Alice's quantum state $\rho_{A^n}=\tr \rho_{AB}^{\otimes n}$ when assisted by the quantum helper Bob.

We now proceed to formally define the coding procedure. We define an $(n,\epsilon)$ code for fully quantum source compression with a quantum helper to consist of the following:
\begin{itemize}
\item Alice's encoding operation $\cE_A: T_A A^n \to A_1 M$ so that the c-q state $\sigma_{A_1M}=\cE_A(\rho_{A^n}\otimes \tau_{T_A})$;
\item Bob's encoding operation $\cE_B: T_BB^n \to L$ so that $\sigma_L=\cE_B(\rho_{B^n}\otimes\tau_{T_B})$ where $|L|=2^{nR_2}$;
\item  Charlie's decoding operation $\cD:M LT_{A}' T_B'\to C_1\widehat{A}^n \widehat{L} \widehat{T}_B^\prime$ that produces 
\[
\omega_{A_1C_1\widehat{A}^n \widehat{L}\widehat{T}_B^\prime}=I_{A_1}\otimes \cD(\sigma_{A_1MLT_A'T_B'})
\]
where
\[
\sigma_{A_1MLT_A'T_B'} = \cE_A\otimes\cE_B(\rho_{AB}^{\otimes n}\otimes \Phi_{T_AT_A'}\otimes \Phi_{T_BT_B'});
\]
\end{itemize}
so that the final state $\omega_{A_1C_1\widehat{A}^n \widehat{L}\widehat{T}_B^\prime}$ satisfies
\begin{align}\label{eq_cond1}
\| \omega_{A_1C_1\widehat{A}^n \widehat{L}\widehat{T}_B^\prime}- \Phi_{A_1C_1}\otimes\rho_{{A}^n L T_B^\prime}\|_1\leq \epsilon,
\end{align}
where $\ket{\Phi_{A_1C_1}}$ is a maximally entangled state.

Note that Bob's message $L$ is quantum, and its size is limited by $2^{nR_2}$.  On the other hand, Alice's message $M$ is classical so that classical communication from Alice to Charlie is considered as a free resource.  As a result, we only evaluate the net consumption of the shared entanglement, i.e.,
$R_1=\log|T_A|-\log|A_1|$.  A rate pair $(R_1,R_2)$ is said to be \emph{achievable} if for any $\epsilon,\delta>0$ and all sufficiently large $n$, there exists an $(n,\epsilon)$ code with rates $R_1+\delta$ and $R_2+\delta$. The rate region is then defined as the collection of all achievable rate pairs. Our main result is the following theorem. 

\begin{theorem}\label{thm_main}
Given is a bipartite quantum source $\rho_{AB}=\tr_R \psi_{ABR}$. The optimal rate region for lossless source coding of $A$ with a quantum helper $B$ is the set of rate pairs $(R_1,R_2)$ such that 
\begin{eqnarray}
R_1 &\geq & H(A|C)_\phi \\
R_2 &\geq &\frac{1}{2} I(RA;C)_\phi. 
\end{eqnarray}
The state $\phi_{ACER}$ resulting from Bob's application of some CPTP map $\cE_{B\to C}$ is 
\begin{equation}\label{eq_state01}
\ket{\phi_{ACER}}= I_{RA}\otimes U^\cE_{B\to CE}\ket{\psi_{ABR}^\rho}.
\end{equation}
\end{theorem}

\subsection{Direct part}

The direct coding theorem uses the channel simulation method. 
For any local channel ${\cal E}_{B \to C}$ performed by the quantum helper $B$ on his half of bipartite state $\rho_{AB}$, it can be simulated by the decoder using the quantum reverse Shannon theorem (QRST) (Theorem~\ref{thm_QRST}):
\begin{align}
\frac{1}{2} I(RA; C)_\phi[q \to q] + \frac{1}{2} I(E; C)_\phi[qq] \ge \langle {\cal E}: \rho_B \rangle,
\end{align}
where 
\[
\ket{\phi_{ACER}}= I_{RA}\otimes U^\cE_{B\to CE}\ket{\psi_{ABR}^\rho}.
\]
In other words, by using the pre-shared entanglement between the helper and the decoder with rate $\frac{1}{2} I(E;C)_\phi$ and sending quantum message from the helper to the decoder with rate $\frac{1}{2} I(RA; C)_\phi$, the decoder can simulate the quantum state ${\cal E}(\rho_B)$ locally with error goes to zero in the asymptotic sense.

Alice's coding: Once the decoder has the system $C$, then Alice and the decoder start the state merging protocol, using the pre-shared entanglement with rate $H(A|C)_\phi$.

\subsection{Converse part}

Here, we refer to Figure~\ref{fig:QSCQH} for corresponding labels used in the converse proof. For any quantum source compression with a quantum helper that produces a state $\omega_{A_1C_1\widehat{A}^n \widehat{L}\widehat{T}_B^\prime}$, it must satisfy (\ref{eq_cond1}). Thus, we can bound $R_1=\log|T_A|-\log|A_1|$  following steps in the converse proof of the state merging protocol \cite{Horodecki:2006hl} and have
\begin{align}
R_1 &\stackrel{>}{\sim} H(A^n|LT_B') \\
&= \sum_{i=1}^n H(A_i |LT_B' A_{<i}) \\
&= \sum_{i=1}^n H(A_i | U_i) \\
&= n H(A_T | U_TT) \\
&= n H(A | C),
\end{align}  
where we set $U_i := (L,T_B', A_{<i})$ and in the last equality, we relabel $A=A_T$ and $C:=(U_T,T)$.

To bound the quantum communication rate $R_2=\log|L|$, we follow steps in the converse proof of the entanglement-assisted quantum rate-distortion theorem (see equation (21) in \cite{Wilde:2013hp}):
\begin{align}
2 R_2 &\ge I(LT_B' ; R^n A^n) \\
&= \sum_{i=1}^n  I(LT_B'; R_iA_i|R_{<i}A_{<i}) \\
&= \sum_{i=1}^n [ I(LT_B'R_{<i}A_{<i}; R_iA_i)- I(R_{<i}A_{<i} ; R_iA_i)] \\
&= \sum_{i=1}^n I(LT_B'R_{<i}A_{<i}; R_iA_i) \\
&\ge \sum_{i=1}^n I(LT_B' A_{<i}; R_iA_i) \\
&= \sum_{i=1}^n I(U_i ; R_iA_i) \\
&= n I(U_T ; R_T A_T|T) \\
&= n I(U_TT; R_TA_T) \\
&= n I(C; RA).
\end{align}
Note that $U_i$ can be generated from $B_i$ via Bob's local CPTP.
In fact, Bob can first append the maximally entangled states $(T_B,T_B')$, systems $(A_{<i},B_{<i})$, and $B_{>i}$. Then, he can perform ${\cal E}_B$, and get $U_i = (L,T_B', A_{<i})$.

\section{Discussion}
The rate region in our main result, Theorem~\ref{thm_main}, bears a close resemblance to its classical counterpart. Our result also shows a helper's strategy of simply compressing the side information $H(C)_\phi$ and sending it to the decoder is sub-optimal with entanglement assistance. Recall the following identity:
\[
H(C)_\phi = \frac{1}{2} I(C;E)_\phi + \frac{1}{2} I(C;RA)_\phi,
\]
where the state $\ket{\phi_{ACER}}$ is given in (\ref{eq_state01}).
The QRST protocol allows us to cleverly divide the amount of quantum communication required for lossless transmission of system $C$ to the decoder into pre-shared entanglement with rate $\frac{1}{2}I(C;E)_\phi$ and quantum communication with rate $ \frac{1}{2} I(C;RA)_\phi$.

We will like to point out that the definition of the fully quantum source compression with a quantum helper requires to explicitly include additional quantum systems $L T_{B}'$ (see Eq.~(\ref{eq_cond1})) for a technical purpose. The reason behind this is because when the quantum state merging is performed, the target systems to which the quantum state is merged needs to be specified. We believe that the inclusion of these additional systems in the definition is inevitable, and it signals a fundamental difference between the fully quantum source compression with a quantum helper and its classical counterpart.  

It should be also noted that our problem formulation is somewhat asymmetric in the sense that
the efficiency of Bob's coding is evaluated by qubits while the efficiency of Alice's coding
is evaluated by the net consumption of the shared entanglement used by Alice. In a future work, it is desirable to consider the amount 
of Alice's communication and the net consumption of the shared entanglement used by Bob as well.

Note that it is possible to replace the state merging protocol with the FQSW protocol, and derive an alternative theorem for quantum source compression with a quantum helper.  It is also possible to consider the same problem without entanglement assistance between the helper and the decoder. These extensions will be treated in the future.

Finally, in the classical source coding with a helper problem, it is possible to bound the dimension for the helper's output system. However, such a result is not unknown to be possible in the quantum regime. 

\section*{Acknowledgements}
MH is supported by an ARC Future Fellowship under Grant FT140100574. 
SW was supported in part by JSPS Postdoctoral Fellowships for Research Abroad.

\end{document}